\documentclass[amsmath,amssymb,floatfix,aps,pra,twocolumn,superscriptaddress]{revtex4-2}

\usepackage{xurl}
\usepackage{graphicx}%
\usepackage{dcolumn}%
\usepackage{layouts}
\usepackage{floatrow}
\usepackage[caption=false,font=footnotesize]{subfig}
\usepackage{braket}

\newcommand{\Gate}[1]{\textsc{#1}}

\newcommand{\zgate}{\Gate{z}}
\newcommand{\ygate}{\Gate{y}}
\newcommand{\xgate}{\Gate{x}}

\newcommand{\xb}{\mathbf{x}}
\usepackage{bm} %

\begin{document}

\title{Importance of Kernel Bandwidth in Quantum Machine Learning}

\author{Ruslan Shaydulin}
\email[]{ruslan.shaydulin@jpmchase.com}
\affiliation{Global Technology Applied Research, JPMorgan Chase, New York, NY 10017 USA}
\affiliation{Mathematics and Computer Science Division, Argonne National Laboratory, Lemont, IL 60439 USA}

\author{Stefan M.\ Wild}
\affiliation{Mathematics and Computer Science Division, Argonne National Laboratory, Lemont, IL 60439 USA}

\date{\today}

\begin{abstract}
Quantum kernel methods are considered a promising avenue for applying quantum computers to machine learning problems. Identifying hyperparameters controlling the inductive bias of quantum machine learning models is expected to be crucial given the central role hyperparameters play in determining the performance of classical machine learning methods.   
In this work we introduce the hyperparameter controlling the bandwidth of a quantum kernel and show that it controls the expressivity of the resulting model. We use extensive  numerical  experiments with multiple quantum kernels and classical datasets to show consistent change in the model behavior from underfitting (bandwidth too large) to overfitting (bandwidth too small), with optimal generalization in between.  We draw a connection between the bandwidth of classical and quantum kernels and show analogous behavior in both cases. Furthermore, we show that optimizing the bandwidth can help mitigate the exponential decay of kernel values with qubit count, which is the cause behind recent observations that the performance of quantum kernel methods decreases with qubit count. We reproduce these negative results and show that if the kernel bandwidth is optimized, the performance instead improves with growing qubit count and becomes competitive with the best classical methods.
\end{abstract}

\maketitle

\section*{Introduction}

Algorithms designed for quantum computers are theoretically able to provide exponential speedups over the best known classical algorithms for certain problems, most famously integer factoring~\cite{shoralg}. Recent advances in quantum computing hardware open the possibility of realizing this potential.
Urgently  needed, however, is development of novel algorithms that effectively leverage the power of quantum computation. A particularly promising problem domain is machine learning, owing to the ubiquity of machine learning problems in science and technology. 

A number of approaches have been proposed for applying quantum computers to machine learning problems. In this work we focus on a subset of such approaches that can be reformulated as kernel methods. The central idea of kernel methods is to embed the data into a feature space (typically, of dimension higher than the input data) in which it becomes easier to analyze. The analysis is then performed by using solely the values of similarities (kernel values) between the representations of the data points in feature space. Most supervised machine learning methods (quantum or classical) can be equivalently reformulated as kernel methods with an appropriately defined (potentially, quantum) kernel~\cite{schuld2021supervised,pmlr-v80-belkin18a}. 
Quantum kernel methods~\cite{Schuld2019,Havlek2019}
have been shown to theoretically provide speedups over classical methods~\cite{Liu2021,Huang2021}. In quantum kernel methods, datapoints are mapped to quantum states using a quantum feature map, and the value of the kernel between two datapoints is given by some similarity measure (such as fidelity) of the corresponding quantum states.  The power of quantum kernel methods comes from being able to process the data using an exponentially sized Hilbert space, performing computations that are hard classically even for sizes attainable on currently available quantum hardware~\cite{Arute2019}. 

Recently, the exponential dimensionality of the space into which the classical data is being mapped has been identified as a potential obstacle to achieving quantum advantage using kernel methods. The fidelity of two random quantum states decreases exponentially with the number of qubits (typically set to be equal to the number of dimensions of the input data), leading to the exponential vanishing of kernel values and making learning impossible~\cite{Banchi2021}.
This behavior is analogous to the curse of dimensionality in classical kernel methods~\cite{Bengio2005curse} and not specific to the choice of fidelity as the similarity measure. One recently proposed approach to overcome this limitation is controlling the inductive bias of the quantum kernel methods by projecting the quantum state into a lower-dimensional subspace~\cite{kubler2021inductive,Huang2021}. In general, however, additional information is required to appropriately choose the projection~\cite{kubler2021inductive}.

These no-go results~\cite{kubler2021inductive,Huang2021} depend crucially on the quantum feature map being fixed as the number of qubits grows. However, this overlooks the important role hyperparameters play in machine learning. Hyperparameter tuning is known to be central to the good performance of classical machine learning models in general and classical kernel methods in particular~\cite{pmlr-v80-belkin18a}. One such hyperparameter is kernel bandwidth, which is known to affect the performance of methods such as support vector machines (SVMs) and is routinely optimized when SVMs are used in practice. Recently, attempts have been made to introduce similar hyperparameters into the quantum models~\cite{2001.03622,2105.03406,Schuld2021}. At the same time, no evidence has been shown that these hyperparameters enable the quantum model to maintain performance as the number of qubits grows, which is a prerequisite for quantum advantage.

In this work we identify quantum kernel bandwidth as a centrally important hyperparameter for quantum kernel methods. We demonstrate that varying it controls the expressiveness of the model, with high bandwidth leading to underfitting and low bandwidth to overfitting. We show that for the quantum feature maps considered, the bandwidth can be controlled by rescaling the datapoints. We draw an analogy with the bandwidth of classical radial basis function (RBF) kernels and show that varying bandwidth affects expressiveness of the model in a similar way in both quantum and classical cases. As a consequence of the effect of the bandwidth on generalization, optimizing the bandwidth can improve the performance from being comparable to a random guess to being competitive with the best classical methods. We reproduce recent numerical results used to support an argument against quantum kernels~\cite{Huang2021}. We show how changing the value of the kernel bandwidth can lead to significantly better performance than the results in~\cite{Huang2021}. %
Moreover, we observe that when the bandwidth is optimized, the performance of quantum kernel methods improves with qubit count, which is the opposite scaling behavior from the one documented in~\cite{Huang2021}. We provide extensive numerical evidence of the importance of the bandwidth and its effect on model generalization and performance with varying qubit count, performing simulations with multiple quantum feature maps and datasets using up to 26 qubits. We show that hardware limitations, such as finite precision of controls and the variance introduced by sampling, are not obstacles to achieving the observed performance.

\section*{Results}

\subsection*{Setup}

We begin by formalizing our notions of quantum kernels, quantum feature maps, and machine learning problems, as well as establishing notation. We consider the problem of supervised learning, specifically the task of classification, although we expect our conclusions to apply more generally. Given a training dataset of $N$ pairs $\{(\mathbf{x}_{i},y_i)\}_{i=1}^N$, where $\mathbf{x}_{i}\in \mathbb{R}^d$ is a datapoint and $y_i\in \{0,1\}$ is a binary label produced by some unknown map $m:\mathbb{R}^d\rightarrow \{0,1\}$, the goal is to learn a map from datapoints to labels that agrees with the true map $m$ with high probability on an unseen test set. 

We consider quantum models where a datapoint $\mathbf{x}_{i}$ is encoded in a quantum state $\ket{\mathbf{x}_{i}}$ by a parameterized unitary: $\ket{\mathbf{x}_{i}} = U_{\text{enc}}(\mathbf{x}_{i})\ket{0}$. We will refer to the unitary $U_{\text{enc}}$ as a quantum feature map. A kernel matrix $K$ is obtained by computing $K_{ij} = k(\mathbf{x}_{i},\mathbf{x}_j) = |\braket{\mathbf{x}_{i}|\mathbf{x}_{j}}|^2$ (the quantum kernel) for all pairs of datapoints. This value can be computed on a quantum computer by measuring the value of observable $\ket{0}\bra{0}$ on the state $U_{\text{enc}}^\dagger(\mathbf{x}_{i})U_{\text{enc}}(\mathbf{x}_{j})\ket{0}$. This kernel matrix is then used inside an SVM or other kernel methods~\cite{scholkopf2002learning, Hastie2009}. A quantum feature map and a classical kernel method such as an SVM fully define a quantum kernel method. %

 In this paper we use the quantum kernel matrix in a support vector classifier (SVC). SVC seeks an optimal separating hyperplane between two (potentially nonseparable) classes in the feature space. An optimal hyperplane is one that maximizes the distance (margin) to the closest point from either class~\cite{vapnik1995nature}. The hyperplane is found by determining coefficients $\bm{\alpha}\in \mathbb{R}^N$ that maximize the objective function
 \begin{equation}
     \sum_{i=1}^{N} \alpha_{i}-\frac{1}{2} \sum_{i=1}^{N} \sum_{j=1}^{N} \alpha_{i} \alpha_{j} y_{i} y_{j} k(\mathbf{x}_{i}, \mathbf{x}_{j}),
     \label{eq:SVC_dual_obj}
 \end{equation}
 subject to $\sum_{i=1}^{N} \alpha_{i} y_{i}=0$ and $0\leq\alpha_i\leq C$, $i=1,\ldots,N$. The penalty term $C>0$ controls the trade-off between the goals of minimizing the training error and maximizing the margin: smaller values of $C$ lead to larger margin at the cost of higher training error.

\floatsetup[figure]{style=plain,subcapbesideposition=top}
\begin{figure*}
    \centering
    \sidesubfloat[]{\includegraphics[width=0.6\textwidth]{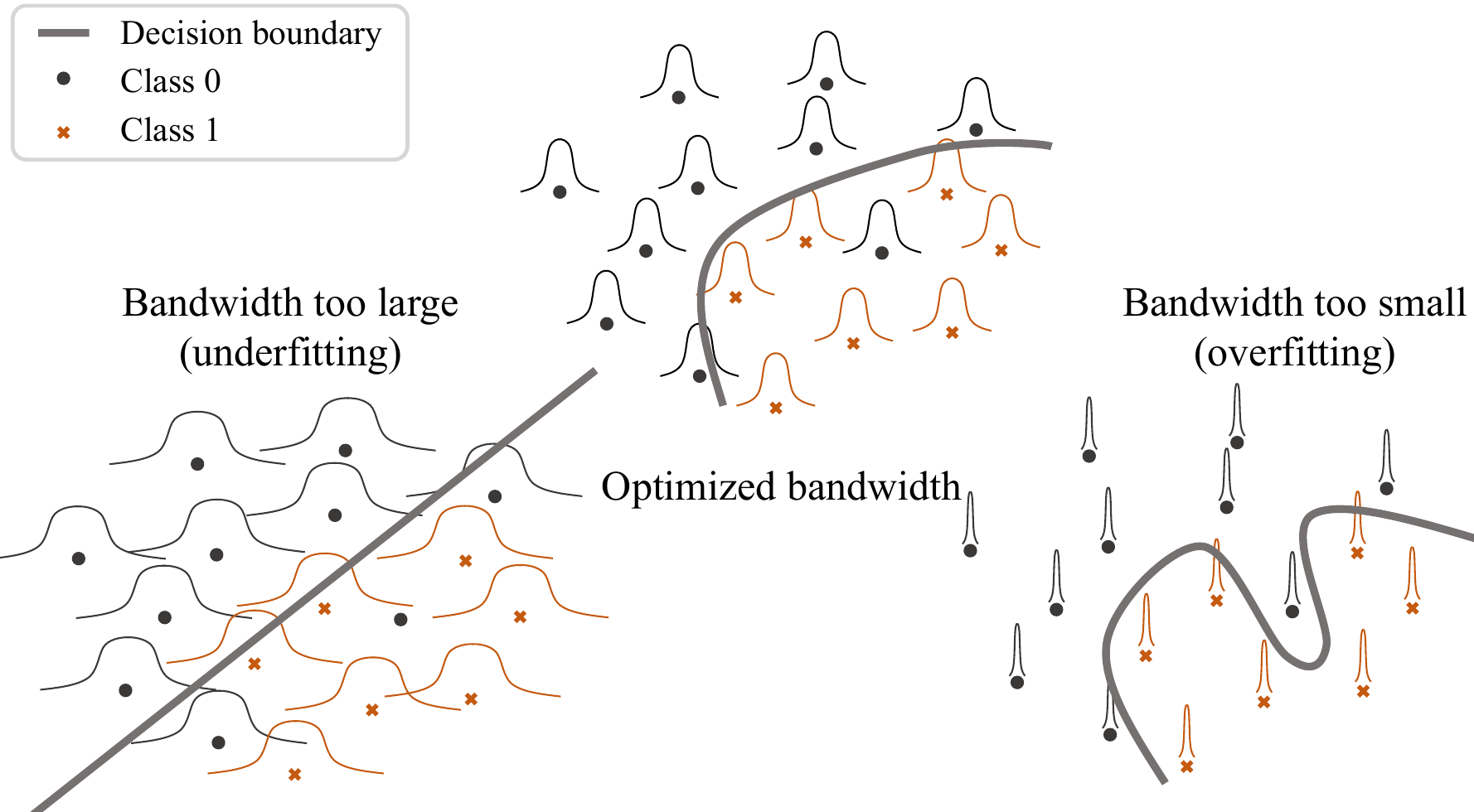}\label{fig:overview}}
    \hspace{0.2in}
    \sidesubfloat[]{\includegraphics[width=0.3\textwidth, trim={0.3in 0 0 0}]{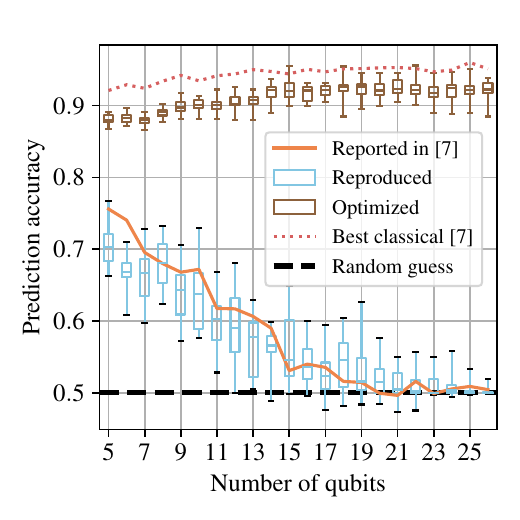}\label{fig:reproduce_google}}
    \caption{\textbf{(a)} Schematic overview of the effect of bandwidth on kernel method performance. Some bandwidth optimization is typically required to avoid under- or overfitting. \textbf{(b)} Prediction accuracy of the quantum kernel method with Hamiltonian evolution feature map as a function of number of qubits. 
     ``Reported'' and ``Reproduced'' results have $t=\frac{d}{3}$, where $d$ is the dimensionality of the datapoint. ``Optimized'' presents performance of the kernel method with the same feature map and $t=0.05$. The performance of our quantum kernel with optimized bandwidth approaches that of the best classical  methods reported in \cite{Huang2021}. Note that from the construction given in Ref.~\cite{Huang2021} (see equation~\ref{eq:HamEvo}), an $n$-dimensional datapoint is embedded into an $(n+1)$-qubit state.%
     }
\end{figure*}
\subsection*{Motivating example}

The recent results on quantum kernel methods do not explicitly study the choice of hyperparameters in quantum feature maps. As a motivating example, we consider the paper~\cite{Huang2021}, which includes numerical experiments on the performance of the quantum kernel method with the feature map given by
\begin{equation}
\left|\mathbf{x}_{i}\right\rangle = \left(\prod_{j=1}^{d} \exp \left(-\mathrm{i} \frac{t}{T} x_{i j}H^{\xgate\ygate\zgate}_j\right)\right)^{T} \bigotimes_{j=1}^{n+1}\left|\psi_{j}\right\rangle, 
    \label{eq:HamEvo}
\end{equation}
where
\begin{equation}
   H^{\xgate\ygate\zgate}_j  = \xgate_{j} \xgate_{j+1}+\ygate_{j} \ygate_{j+1}+\zgate_{j} \zgate_{j+1}. 
\end{equation}
Here $\xgate_j$, $\ygate_j$, and $\zgate_j$ are the Pauli operators acting on qubit $j$, and $\ket{\psi_j}$ is a Haar-random single-qubit state. The Haar-random state $\ket{\psi_j}$ is sampled and fixed for each qubit. This feature map represents a $d$-dimensional datapoint as a $d+1$-qubit quantum state~\cite{codeHuang2021}. %
We refer to this feature map as the Hamiltonian evolution feature map.

The feature map given by equation~\ref{eq:HamEvo} has two hyperparameters: the total evolution time $t$ (equivalent to the scaling of the inputs $\mathbf{x}_i \leftarrow t\mathbf{x}_i$) and the  number of Trotter steps $T$. Huang  et  al.~\cite{Huang2021} set $t=\frac{d}{3}$ and $T=20$ and reported prediction accuracy on the test set of the quantum kernel method (SVC) with this feature map on the Fashion-MNIST (\texttt{fmnist}) dataset~\cite{xiao2017fmnist}. Each datapoint is reduced to the specified number of dimensions by using principal component analysis (PCA). In Fig.~\ref{fig:reproduce_google} we present the results reported in ~\cite{Huang2021} and the results we reproduced independently. The details of the SVC implementation and data preprocessing are given in the Methods section. Our reproduced results are presented as boxplots over 20 choices of the  Haar-random initial state fixed across all qubit counts, with the box showing the middle $50\%$ quantile range and whiskers showing minimum and maximum. 
The reported and reproduced results agree well, and both show that the performance of the quantum kernel method decreases with increasing qubit count, approaching the performance of a random guess (i.e., 0.5 accuracy) at about 20 qubits.

While Huang  et al.~\cite{Huang2021} performed extensive hyperparameter optimization of both the SVC that uses the quantum kernel and the classical methods with which the quantum kernel method is compared, no hyperparameter optimization is reported for the quantum feature map itself. In fact, choosing $t=0.05$ and keeping everything else fixed lead to much better performance, approaching the performance of the  best classical machine learning methods (see Fig.~\ref{fig:reproduce_google}). The reason is that the parameter $t$ controls the bandwidth of the quantum kernel. Large values of $t$ lead to the kernel's being too ``narrow'' (small bandwidth), making learning impossible. We discuss the relationship between the scaling factor, quantum kernel bandwidth, and model performance in detail below.

In addition to providing improved performance, this hyperparameter choice leads to an opposing conclusion about the scaling of the performance with qubit count: in this case, the performance increases with qubit count until plateauing at about 14 qubits. The improvement in performance can be partially explained by observing that with larger qubit count, more components are kept in the PCA-based reduction, thereby giving the model access to a larger part of the explanation of variance in the data. Unlike the previous results~\cite{kubler2021inductive,Huang2021}, we do not observe a large drop in performance for larger qubit counts. This example highlights the importance of hyperparameter selection in determining the effectiveness of quantum kernel methods. 

\subsection*{Importance of hyperparameters in quantum machine learning}

We now show that the ability of hyperparameter choice to make or break performance of a quantum kernel method is not limited to the above example. To this  end, we perform a systematic study of how bandwidth affects the prediction accuracy of quantum kernel methods. In addition to the Hamiltonian evolution feature map given in Equation~\ref{eq:HamEvo}, we consider a feature map inspired by the instantaneous quantum polynomial (IQP) circuits used in \cite{Havlek2019,Huang2021}:
\begin{equation}
    \left|\mathbf{x}_{i}\right\rangle=U_{Z}\left(\mathbf{x}_{i}\right) H^{\otimes d} U_{Z}\left(\mathbf{x}_{i}\right) H^{\otimes d}\left|0^{d}\right\rangle,
    \label{eq:IQP_feature_map}
\end{equation}
where
\begin{equation}
    U_{Z}\left(\mathbf{x}_{i}\right)=\exp \left(\sum_{j=1}^{d} \lambda x_{i j} Z_{j}+\sum_{j=1}^{d} \sum_{j^{\prime}=1}^{d}  \lambda^2 x_{i j} x_{i j^{\prime}} Z_{j} Z_{j^{\prime}}\right).
    \label{eq:IQP_operator}
\end{equation}

\begin{figure*}[ht]
\centering %
\sidesubfloat[]{\includegraphics[width=0.66\textwidth, trim=1.2in 0 0.8in 0.1in]{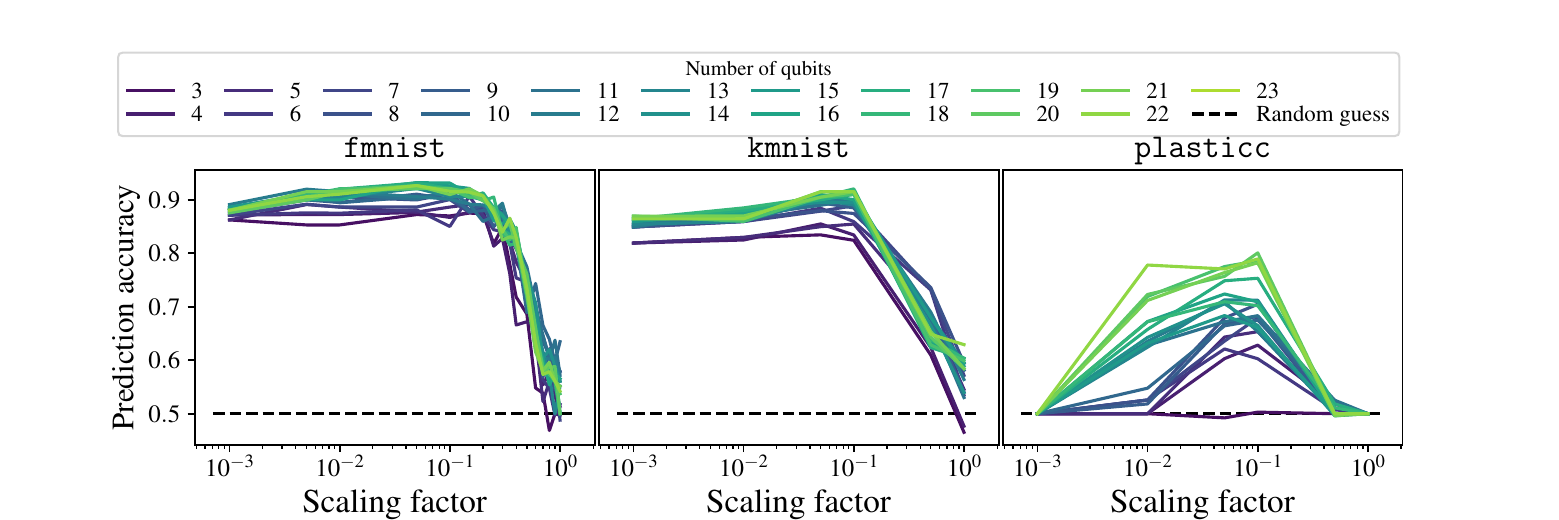}\label{fig:scaling_iqp}}
\sidesubfloat[]{\includegraphics[width=0.26\textwidth, trim=0.1in 0 0 0]{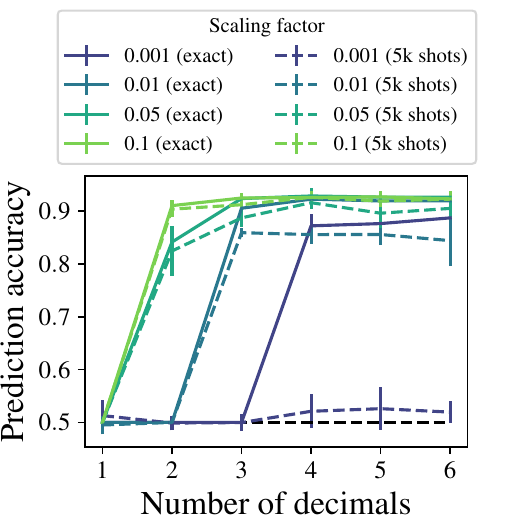}\label{fig:finite_prec_sampl}} \\
\sidesubfloat[]{\includegraphics[width=0.66\textwidth, trim=1.2in 0 0.8in 0.75in]{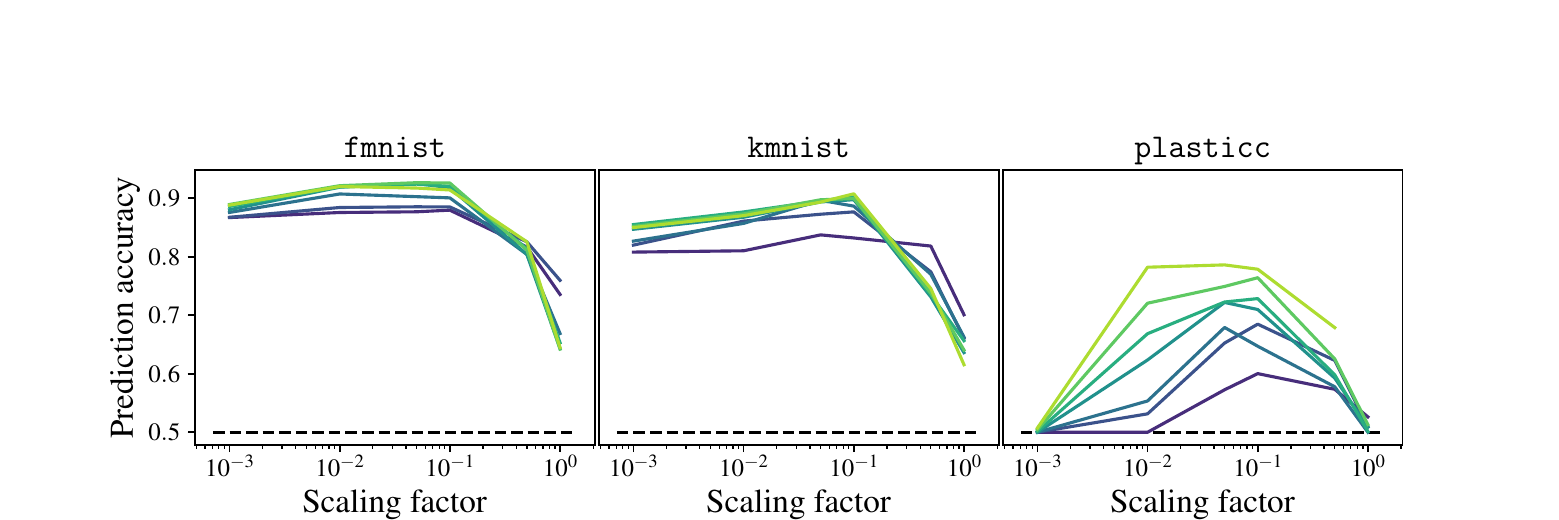}\label{fig:scaling_HamEvo}}
\sidesubfloat[]{\includegraphics[width=0.26\textwidth, trim=0.1in 0in 0in 0.25in]{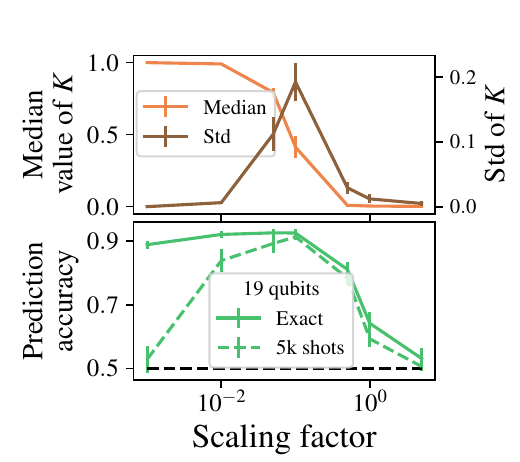}\label{fig:full_prec_sampl}}
\caption{Prediction accuracy of the quantum kernel method with IQP \textbf{(a)} and Hamiltonian evolution \textbf{(c)} feature maps and as a function of the scaling factor. \textbf{(b)} Effects of finite precision of controls on performance of the quantum kernel method. Error bars show one standard deviation. \textbf{(d)} Effects of finite sampling on performance of the quantum kernel method.
Top: median value and standard deviation of the non-diagonal entries of the kernel matrix $K$ computed from full-precision quantum state. Bottom: test accuracy for kernel matrix computed from full-precision quantum state and from 5,000 samples as a function of scaling factor applied to the data.}
\label{fig:stream}
\end{figure*}

We endow this feature map with a hyperparameter $\lambda>0$ analogous to the total evolution time in the Hamiltonian evolution feature map. Since both hyperparameters are equivalent to a rescaling of the input data, we refer to them as ``scaling factors.'' %
The scaling factor $\lambda$ can be subsumed into the definition of the input data by setting $\mathbf{x}_i \leftarrow \lambda\mathbf{x}_i$. The scaling factor can be understood as a hyperparameter of the quantum feature map since it can always be included in the definition of the feature map if an assumption on the input data distribution is made (e.g., the data has mean zero and standard deviation one). We show below that the scaling factor behaves analogously to bandwidth in classical kernel methods. 

The performance is evaluated on three real-world datasets. In addition to the Fashion-MNIST (\texttt{fmnist}) dataset, we use Kuzushiji-MNIST (\texttt{kmnist}), a dataset containing cursive Japanese characters~\cite{kmnist}, and a cosmological dataset focused on the task of supernova classification (\texttt{plasticc})~\cite{plasticc}. Two of the datasets were used to study quantum kernel methods (\texttt{fmnist} in~\cite{Huang2021} and \texttt{plasticc} in~\cite{peters2021machine}). The datasets are balanced; that is, the number of datapoints with label $0$ is approximately equal to the number of datapoints with label $1$ for both training and testing datasets. All input data is normalized to be centered around zero with the standard deviation of approximately one. A precise description of the experimental setup is given in the Methods section.

We observe that hyperparameter selection significantly affects the performance of quantum kernel methods. We find in particular that the performance of the models varies drastically with the scaling factor (bandwidth). This relationship is presented in Figs.~\ref{fig:scaling_iqp} and  \ref{fig:scaling_HamEvo}. Since no significant effect is observed from varying the number of Trotter steps $T$ in the Hamiltonian evolution feature map, here and for the remainder of the paper we set $T=40$; results with other values are available in the Supplementary Information. We observe a clear trend in optimal values of the scaling factor, with larger values leading to overfitting and smaller values to underfitting. This relationship is schematically presented in Fig.~\ref{fig:overview}. Counterintuitively, the scaling factor of one (corresponding to the data having standard deviation of approximately one) is a bad choice for all the datasets and feature maps considered.

\begin{figure*}[t]
    \centering
    \sidesubfloat[]{
    \includegraphics[width=0.22\textwidth, trim={0.4in 0 0 0}]{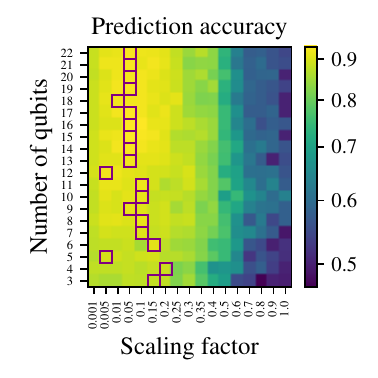}\label{fig:IQP_2d_score}}
    \sidesubfloat[]{
    \includegraphics[width=0.22\textwidth, trim={0.4in 0 0 0}]{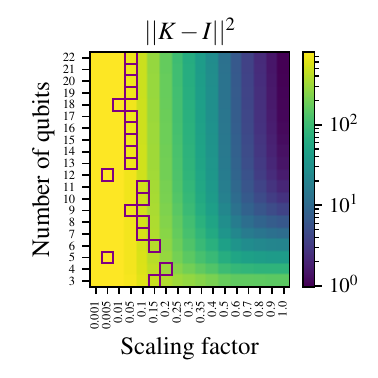}\label{fig:IQP_2d_norm}}
    \sidesubfloat[]{
    \includegraphics[width=0.2\textwidth, trim={0.3in 0 0 0.5in}]{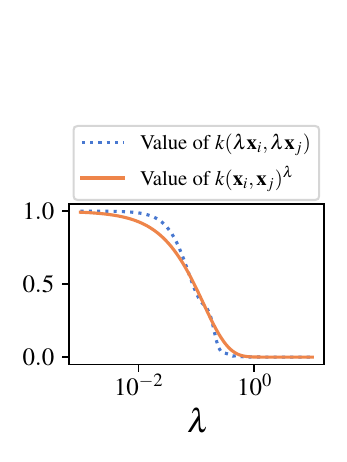}\label{fig:IQP_kern_value_vary_lambda}}
    \sidesubfloat[]{
    \includegraphics[width=0.23\textwidth, trim={0.3in 0 0 0}]{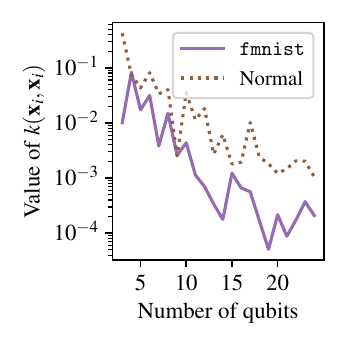}\label{fig:IQP_kern_value_vary_nqubits}}
    \caption{Scaling of the kernel with the IQP quantum feature map. \textbf{(a)} Prediction accuracy as a function of number of qubits and the scaling factor $\lambda$. For each qubit count, the value of the scaling factor corresponding to the highest accuracy is highlighted with a purple square. The trend in scaling factor values corresponding to the highest accuracy is evident. \textbf{(b)} Kernel matrix entries as a function of number of qubits and the scaling factor $\lambda$. If $\|K-I\|^2$ is small, non-diagonal kernel matrix entries are small (i.e., $K$ approaches the identity).  The value $\|K-I\|^2$ is largest when all kernel matrix entries are close to one.  \textbf{(c)} The scaling of the kernel as a function of the scaling factor $\lambda$. The points  $\xb_i$ and $\xb_j$ are taken from the \texttt{fmnist} dataset; the number of qubits is 20. The kernel scales exponentially with $\lambda$. \textbf{(d)} Value of the kernel of two points as a function of the number of qubits. Exponential decay in the value of the kernel entry with the number of qubits is clearly visible.}
    \label{fig:kern_scale}
\end{figure*}

We can understand the importance of the scaling factor by considering the relationship between qubit count, scaling factor, and the values of the kernel matrix. 
The performance of SVC is sensitive to kernel bandwidth. If the kernel bandwidth is too small (kernel is ``narrow''), then the kernel matrix will be close to identity; and if the bandwidth is too large (kernel is ``wide''), all kernel matrix entries will be close to one. In both scenarios learning is impossible, which is why kernel bandwidth is routinely optimized in classical kernel methods. Scaling factor is analogous to kernel bandwidth for quantum kernels. When the scaling factor is large, the kernel matrix approaches identity as the qubit count grows. Correspondingly, when the scaling factor is small, kernel matrix entries all approach one. This relationship is shown in Fig.~\ref{fig:IQP_2d_norm}. Both of those extremes lead to suboptimal performance (see Fig.~\ref{fig:IQP_2d_score}); the goal of hyperparameter optimization is therefore to find the ``Goldilocks'' value of the scaling factor. We discuss the analogy with classical kernel bandwidth and the effect on model generalization in detail below by comparing the behavior of quantum kernels with that of the classical RBF kernel. Previously, in~\cite{peters2021machine}  rescaling of the data was shown to lead to an improvement in performance of SVC with a different feature map applied to the \texttt{plasticc} dataset.

Let us now consider the impact of scaling factor and qubit count on the values of the kernel matrix. First, note that if the data is rescaled by $\lambda$, then the operator (\ref{eq:IQP_operator}) becomes
\begin{align*}
U_{Z}\left(\lambda\mathbf{x}_{i}\right) =\exp \left(\lambda\sum_{j=1}^{d} x_{i j} Z_{j}+\lambda^2\sum_{j=1}^{d} \sum_{j^{\prime}=1}^{d} x_{i j} x_{i j^{\prime}} Z_{j} Z_{j^{\prime}}\right) \\
= \left(\exp \left(\sum_{j=1}^{d} x_{i j} Z_{j}+\lambda\sum_{j=1}^{d} \sum_{j^{\prime}=1}^{d} x_{i j} x_{i j^{\prime}} Z_{j} Z_{j^{\prime}}\right)\right)^\lambda.    
\end{align*}

Analogous behavior can be easily shown for the Hamiltonian evolution feature map. This highlights the exponential relationship between the scaling factor and the kernel value for the family of kernels considered in this work. Figure~\ref{fig:IQP_kern_value_vary_lambda} numerically elucidates this exponential relationship by plotting $k(\lambda\mathbf{x}_i, \lambda\mathbf{x}_j)$ against $k(\mathbf{x}_i, \mathbf{x}_j)^\lambda$ for two fixed datapoints. Second, note the exponential relationship between the qubit count and the kernel value. As the number of qubits grows, the fidelity of two random states decays exponentially. Correspondingly, the fidelity of the embeddings $\ket{\mathbf{x}_i}$ of the datapoints, equal to the value of the quantum kernel of the datapoints, also decays exponentially. This can be observed in Fig.~\ref{fig:IQP_kern_value_vary_nqubits}, which shows the scaling of the value of the kernel of two points with the number of qubits. The points in Fig.~\ref{fig:IQP_kern_value_vary_nqubits} are sampled from a normal distribution with mean zero and standard deviation of one (``Normal'') or are taken from the \texttt{fmnist} dataset, preprocessed as described in the Methods section, and scaled to have mean zero and standard deviation one. In both cases, the exponential decay in the kernel value with qubit count is clearly visible. These two exponential relationships have important implications since high qubit count is a necessary (but not sufficient) prerequisite for the possibility of quantum advantage. While the kernel value decays exponentially irrespective of the fixed value of the scaling factor, selecting a scaling factor based on the qubit count may mitigate the effects of this decay.

For the datasets and feature maps considered, and the range of qubit counts accessible in classical simulation, we find that choosing one fixed scaling factor leads to good performance. However, we expect that in general the value of scaling factor will vary with qubit count, data properties and the feature map structure. For example, we anticipate that the scaling factor required to maintain good generalization will decrease with qubit count. Initial evidence of this can be glanced from Figs.~\ref{fig:IQP_2d_score}, wherein the optimized value of the scaling factor decreases slightly for higher qubit count. Since obtaining true scaling of the bandwidth parameter would require experiments with the numbers of qubits far beyond those accessible in simulation, a promising future direction is studying of analytically tractable feature maps and data distributions such as those considered in Ref.~\cite{kubler2021inductive}. For such problems it may be possible to derive closed-form scaling of optimal bandwidth parameter, as well as establish the scaling of important properties such as the range of parameter values that lead to good generalization.

\begin{figure}
    \centering
    \includegraphics[width=2.2in]{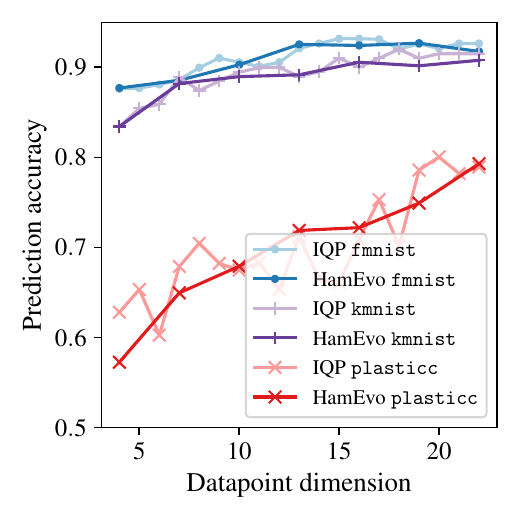}
    \caption{Prediction accuracy with varying datapoint dimension for feature maps with optimized hyperparameters for a given dataset. One value of hyperparameters is chosen for each dataset and is fixed across all datapoint dimensions. For the Hamiltonian evolution feature map, mean accuracy over all initial state seeds is presented. The datapoint dimension is equal to the number of qubits for the IQP feature map and is equal to the $(\text{number of qubits} - 1)$ for the Hamiltonian evolution (``HamEvo'') feature map. For the easier datasets, \texttt{fmnist} and \texttt{kmnist}, the performance stops improving after a certain threshold is reached, whereas for the harder dataset, \texttt{plasticc}, the performance keeps improving up to the largest datapoint dimension considered.}
    \label{fig:performance_vs_nqubits_all}
\end{figure}

As shown in Fig.~\ref{fig:reproduce_google}, the Hamiltonian evolution feature map with optimized scaling factor can achieve accuracy approaching that of the best classical model reported in ~\cite{Huang2021}. The improvement in performance with qubit count, observed in the motivating example, is present for all datasets and qubit counts considered. Figure~\ref{fig:performance_vs_nqubits_all} shows prediction accuracy as a function of datapoint dimension, which is equal to the qubit count in the case of the IQP feature map and is equal to $(\text{Number of qubits} - 1)$ for the Hamiltonian evolution feature map. With the scaling factor fixed to the optimized value, we observe improvement in performance with growing qubit count up to a certain threshold at which the performance plateaus. The value of the threshold depends on the difficulty of the classification task. For \texttt{fmnist}, considered in \cite{Huang2021}, the performance stops improving at about 13 qubits, whereas for the more difficult \texttt{plasticc} dataset, considered in \cite{peters2021machine}, the performance improves for the entire range of qubit counts considered, namely,  up to 22. Similar to the motivating example, the improvement in performance can be partially attributed to the increased dimensionality of the input datapoint, which allows PCA to preserve more data variance. We note, however, that within the regime considered, no deleterious effects of increased dimensionality of the Hilbert space are observed. The cessation in the performance improvement with qubit count for \texttt{fmnist} can be explained by the model saturating the best possible performance on this problem, as evidenced by it matching the performance of the best known classical methods (see Fig.~\ref{fig:reproduce_google}). 

\subsection*{Impact of bandwidth on model generalization and comparison with classical kernels}

In the preceding section we showed that the performance of quantum kernel methods is sensitive to quantum kernel bandwidth (scaling factor). This sensitivity is due to how the kernel bandwidth affects the expressivity of the model, which we now explain. We being by noting that the behavior is completely analogous to the variation in the performance of classical kernel methods with kernel bandwidth. The role of kernel bandwidth and its effect on SVC performance are well known in classical machine learning~\cite{scholkopf2002learning,shawe2004kernel,kung2014kernel}. To illustrate its effect, we consider the RBF kernel given by $k(\xb_i, \xb_j) = \exp(-\gamma \|\xb_i - \xb_j\|^2)$; here the hyperparameter $\gamma$ controls the bandwidth. Analogously to the case of the quantum feature map we considered above, $\gamma$ can be subsumed into the definition of input data as a scaling factor. The kernel matrix $K_{ij} = k(\xb_i, \xb_j)$ is then used in an SVC analogously to the quantum kernel case described above.

\begin{figure*}
    \centering %
    \sidesubfloat[]{
    \includegraphics[width=0.2\textwidth, trim=0.1in 0 0 0]{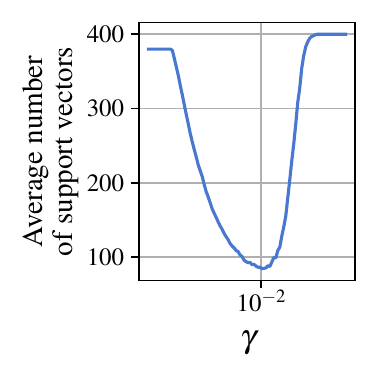}
    \includegraphics[width=0.2\textwidth, trim=0.1in 0 0 0]{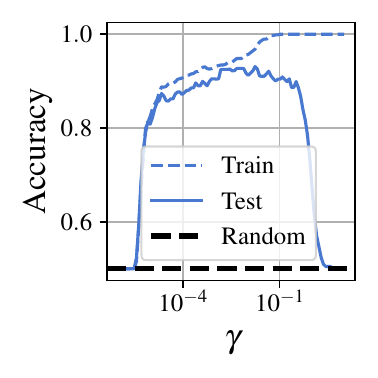}\label{fig:rbf_bandwidth}} 
    \sidesubfloat[]{
    \includegraphics[width=0.2\textwidth, trim=0.1in 0 0 0]{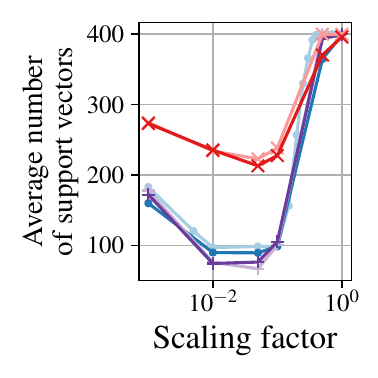}
    \includegraphics[width=0.36\textwidth, trim=0.1in 0 0.9in 0]{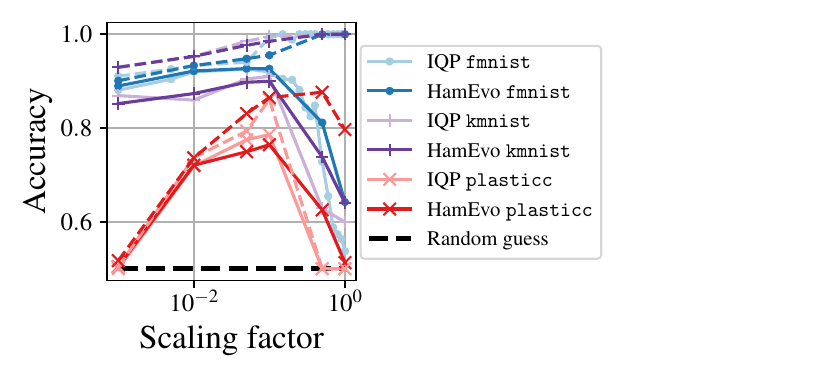}\label{fig:quantum_bandwidth}} \\
    \sidesubfloat[]{\includegraphics[width=0.2\textwidth]{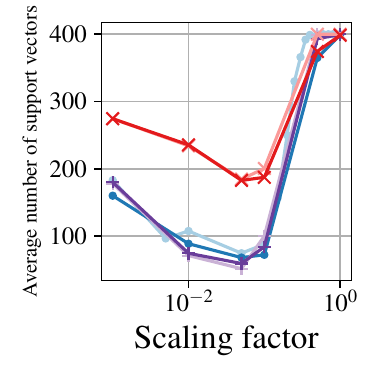}
    \includegraphics[width=0.36\textwidth, trim=0 0 0.89in 0]{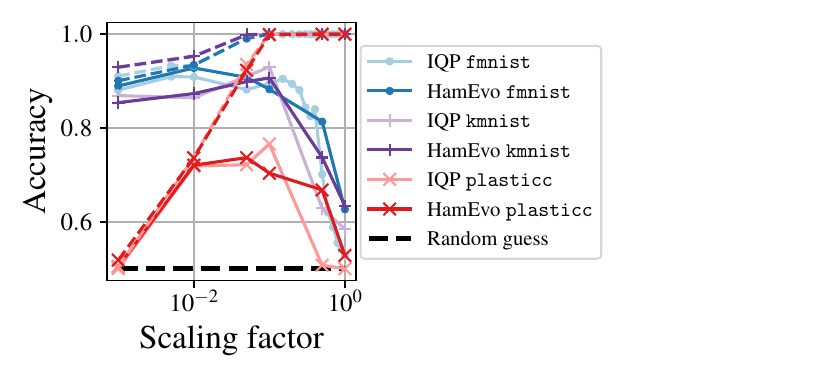}\label{fig:quantum_bandwidth_noCV}}
    \caption{Effects of kernel bandwidth on model generalization on \texttt{fmnist} dataset for \textbf{(a)} classical RBF kernel and  quantum kernels using 19 qubits with \textbf{(b)} and without \textbf{(c)} cross-validation. Dashed and solid lines represent the training and test scores, respectively. For large values of $\gamma$ or the scaling factor, the kernel is narrow, and the models can fit any data, leading to overfitting. For small values, the kernel is wide, and the model is insufficiently expressive, leading to underfitting. The maximum average number of support vectors per class is 400 (i.e., every training point is a support vector).}
    \label{fig:all_bandwidth}
\end{figure*}

If $\gamma$ is large, the kernel is ``narrow,'' and the SVC with such a kernel can fit any labels, which may lead to overfitting. This is analogous to the scaling factor $\lambda$ above, where large scaling factor values lead to overfitting. Figure~\ref{fig:all_bandwidth} shows overfitting resulting from an overly expressive kernel. For large values of $\gamma$ or $\lambda$, the score on the training set goes to one, and the score on the test set decreases to that of random guessing. On the other hand, for sufficiently small values, the kernel is ``wide,'' and the SVC with such kernel is insufficiently expressive, leading to underfitting. The goal is therefore to identify the bandwidth that makes the model sufficiently but not overly expressive.

To illustrate the effect of quantum kernel bandwidth (controlled by the scaling factor) on the expressiveness of the model, we present two sets of experiments in Fig.~\ref{fig:all_bandwidth}. We optimize the penalty term $C$ in SVC either using cross-validation or to directly maximize the training score (see Methods for detailed description). When the penalty term that maximizes the training score is chosen, we observe clear under- and overfitting for small and large values of the scaling factor, respectively. Additional information about the performance of the methods without cross-validation is given in the Supplementary Information. In the preceding sections, however, we presented the results with cross-validation since they show slightly better performance. In this case, for the easier \texttt{fmnist} and \texttt{kmnist} datasets, the behavior is similar. For the harder \texttt{plasticc} dataset, the training score for the largest value of the scaling factor $\lambda$ decreases, thereby indicating the  absence of overfitting. We observe that for the \texttt{plasticc} dataset, for both feature maps the cross-validation hyperparameter optimization leads to the smallest possible value $C$, corresponding to the largest margin at the cost of a  lower training score. For this particular dataset, this is sufficient to prevent overfitting. For the other datasets considered, the hyperparameter optimization does not lead to the largest penalty, suggesting that increasing the penalty would not help overcome overfitting.

\subsection*{Effects of limited control precision and finite sampling}

So far we have demonstrated that with appropriate bandwidth selection, performance of quantum kernel methods need not deteriorate with the growth in qubit count and can approach the performance of the best classical methods. Two major obstacles arise in realizing the observed performance of quantum models in practice: the errors introduced by finite sampling and the limited precision of controls. We now provide numerical evidence that these obstacles do not prevent realizing the good model performance observed above under realistic assumptions on the number of samples used and the control precision.

The first obstacle is that, in practice, the kernel matrix is evaluated by performing measurements of a quantum device, limiting the effective precision with which kernel matrix elements can be evaluated. In general, the errors introduced by the finite sampling statistics need not preclude quantum advantage~\cite{Liu2021}. For small values of the kernel, however, finite sampling may lead to large relative errors~\cite{otten2020GPKernel} and can potentially nontrivially affect prediction performance. Moreover, the variance of the errors in the value of $|\braket{0|\psi}|^2$ introduced by finite sampling differs depending on the choice of state $\ket{\psi}$, necessitating careful numerical investigation. This can be easily observed by comparing the variance between  $\ket{\psi}=\frac{1}{2^n}\sum_{x\in \{0,1\}^n}\ket{x}$ and $\ket{\psi}$ being an n-qubit GHZ state~\cite{varianceDependsOnStateCode}. To numerically evaluate the effects of finite sampling in a realistic setting, we consider the Hamiltonian evolution feature map with the hyperparameters corresponding to the highest prediction accuracy on the \texttt{fmnist} dataset ($n=19$, $T=40$). For each value of the scaling factor, we recompute the 5-by-5 subset of the corresponding kernel matrix using 5,000 shots 10 times, and we use the mean of the standard deviations of the submatrix entries as the standard deviation for i.i.d.\ normal noise with mean zero applied to the full matrix. The noise is applied in a way that preserves the symmetry of the matrix. The resulting matrix may not be positive semi-definite; we take the closest (in Frobenius norm; see, e.g., Ref.~\cite{Higham1988}) positive semi-definite approximation as the kernel matrix for the SVC. 

Figure~\ref{fig:full_prec_sampl} presents the prediction accuracy obtained with the kernel matrix evaluated from full quantum state up to machine precision (``Exact'') and with sampling noise (``5k shots''). We observe poor performance in the cases where the nondiagonal kernel matrix values are all close to zero or all close to one (i.e., kernel very narrow or very wide). However, at the same time, when an appropriate kernel bandwidth (equivalent to scaling the data) is chosen, we observe that good performance is recovered. This is due to bandwidth choice preventing the values of the kernel matrix from concentrating; see Fig.~\ref{fig:full_prec_sampl}, top. This evidence suggests that in some cases judicious choice of bandwidth parameter may reduce the number of samples needed. 

The second obstacle is finite precision of controls available on noisy intermediate-scale quantum (NISQ) hardware, which limits the range of values that can be used as inputs to the quantum feature map. This can potentially limit the scaling of the data. Note that this issue is limited to NISQ devices, since fault-tolerant architectures allow synthesis of single-qubit operations with precision matching that of classical machines~\cite{Wiebe2013}. To numerically evaluate the effects of finite precision, we recompute the kernel matrix values with parameters rounded to no more than a fixed number of decimals. Figure~\ref{fig:finite_prec_sampl} shows that with the scaling factor that maximizes performance in the full-precision case (0.05), three decimals of precision is sufficient to match the performance of the noiseless case. As the scaling factor decreases, the number of decimals required to match the full precision increases. We note that precision of trapped ion hardware currently available on the cloud (three decimals~\cite{ionqprecision}) is sufficient to saturate the noiseless performance of a scaling factor that is five times smaller than the one maximizing performance at 19 qubits, suggesting that the control precision would not be the limiting factor for the qubit counts expected in the near term. 

\section*{Discussion}

Numerical evaluation of promising quantum algorithms is an essential complement to theoretical analysis. When identifying opportunities and no-go results for quantum advantage, it is therefore crucial to consider all aspects of the quantum algorithm that can be used to boost its performance. In this work we identify quantum kernel bandwidth as the key hyperparameter controlling the expressiveness of the model, with excessively small bandwidth leading to overfitting and large bandwidth leading to underfitting. 
 We show that careful kernel bandwidth selection can mitigate the exponential decay of kernel values with qubit count and avoid overfitting and the corresponding drop-off in prediction accuracy. As a pleasant surprise, we discover that for a given dataset a wide range of bandwidth values  work well (this is most clearly visible in Fig.~\ref{fig:all_bandwidth}). This observation suggests that optimizing kernel bandwidth need not be difficult in practice: finding the correct order of magnitude is often sufficient for achieving good performance.
 
 While the bandwidth hyperparameter we introduce is superficially similar to the rescaling of the data used by Schuld et al.~\cite{Schuld2021} to control the expressive power of parameterised quantum circuits, our perspective differs in two important ways. First, we study the kernel method setting, therefore the expressiveness of the model is controlled by the properties of the kernel matrix: roughly speaking, the closer the kernel matrix is to diagonal, the more expressive the model. As can be glanced from Fig.~\ref{fig:all_bandwidth}, one consequence of this is that simply increasing the scaling factor leads to more expressive model -- a drastically different dynamic from the one considered in Ref.~\cite{Schuld2021}. Second, the kernel setting we choose has the attribute that increasing the number of qubits while keeping the feature map fixed implicitly increases the expressiveness of the model. This expressiveness is then modulated by the bandwidth hyperparameter. The setting chosen in Ref.~\cite{Schuld2021} does not have this property, and the authors do not consider explicitly the interplay between the hyperparameters and qubit counts.

Recent proposals to control the inductive bias of quantum kernels by projecting them into a lower-dimensional subspace~\cite{Huang2021,kubler2021inductive} are clearly complementary to the ideas put forward in this work since they attempt to overcome similar challenges. The projection can be straightforwardly combined with bandwidth optimization to more precisely modulate the inductive bias of the model. More interestingly, the projection operation can itself be viewed as a hyperparameter in the  absence of sufficient prior knowledge. Such techniques may be required to achieve good performance with qubit counts significantly larger than  accessible for study at the time of writing.

In this paper we  consider only two hyperparameters---scaling factor and the number of Trotter steps in the Hamiltonian evolution feature map. We see a significant effect for only one of them,  the scaling factor, which controls the kernel bandwidth. We anticipate that more elaborate feature maps will be explored with potentially more parameters that can be tuned. Doing so may help bridge the gap between fully trainable quantum embeddings~\cite{2001.03622}, which require expensive training on a quantum computer, and fixed feature maps. Careful consideration of hyperparameters of quantum kernels may enable the tuning of the inductive bias of the model in a way that is infeasible classically, providing a potential path to quantum advantage in machine learning.

\section*{Methods}

\paragraph*{Data preprocessing}

We preprocess the input data in the following way. For \texttt{fmnist}~\cite{xiao2017fmnist}, we follow the preprocessing pipeline of~\cite{Huang2021}, which results in the two classes of dresses (class 0) and shirts (class 3). For \texttt{kmnist}~\cite{kmnist}, we arbitrarily choose class 1 and class 4. For \texttt{plasticc}~\cite{plasticc}, we use the dataset preprocessed as described in~\cite{peters2021machine}, provided by the authors. For all datasets, each datapoint is rescaled such that the mean value is zero and the standard deviation is one. Then, each datapoint (28-by-28 pixel images for \texttt{fmnist} and \texttt{kmnist} and 67-dimensional vector for \texttt{plasticc}) is reduced to the required number of dimensions by using the scikit-learn~\cite{scikit-learn} implementation of PCA. We use 800 randomly selected points as the training set and 200 randomly selected points as the test set. The \texttt{fmnist} and \texttt{kmnist} datasets provide separate training and test sets, and for \texttt{plasticc} we split the dataset into training and testing parts randomly, as stratified by the value of the label. The random splits result in approximately (though not exactly) balanced problems. The same random samples are used for all quantum feature maps and hyperparameter values.

\paragraph{Kernel methods} The quantum kernels are computed by using Qiskit~\cite{Qiskit}. For each datapoint $\mathbf{x}_{i}$, we use the Qiskit Aer simulator~\cite{Qiskit} to compute the vector of amplitudes describing the quantum state of its representation $\ket{\mathbf{x}_{i}}$. Then, the values of the kernel $k(\mathbf{x}_{i},\mathbf{x}_j) = \left|\braket{\mathbf{x}_{i}|\mathbf{x}_{j}}\right|^2$ are obtained by taking inner products of the corresponding vectors of amplitudes. The resulting kernel matrix is used in the  scikit-learn~\cite{scikit-learn} implementation of the SVC. Following ~\cite{Huang2021}, we perform a  grid search over the following values of the penalty term in Eq.~\ref{eq:SVC_dual_obj}:
\begin{align*}
C\in\{0.006, & 0.015, 0.03, 0.0625, 0.125,0.25, 0.5, 1.0, 2.0,\\
&  4.0, 8.0, 16.0, 32.0, 64.0, 128.0, 256, 512, 1024\}.
\end{align*}

For the results presented in Fig.~\ref{fig:reproduce_google},~\ref{fig:quantum_bandwidth_noCV}, we choose the value $C$ that maximizes the training score. In all other plots, we perform 5-fold cross-validation to choose $C$. In all plots, we report balanced accuracy, defined as the mean recall of the two classes, as the prediction accuracy score. 

For the classical RBF kernel $k(\xb_i, \xb_j) = \exp(-\gamma \|\xb_i - \xb_j\|^2)$, we begin by performing 5-fold cross-validation over the values of $C$ defined above and the following values of $\gamma$:
\begin{align*}
 \gamma \in\{ & 0.25, 0.5, 1.0, 2.0, 3.0, 4.0, 5.0, 20.0, 50.0, 100.0, \\
 & 200.0, 500.0, 1000.0, 5000.0, 10000.0\} /\left(N \operatorname{Var}\left[\mathbf{x}\right]\right),   
\end{align*}
where $\operatorname{Var}\left[\mathbf{x}\right]$ is the variance of the training data and $N=800$ is the number of training points. Then, we  fix the value of $C$ to the optimized value obtained from the cross-validation, and we vary parameter $\gamma$ to obtain Fig.~\ref{fig:rbf_bandwidth}.

\paragraph{Data availability} We release all code used to generate the data and the plots, as well as the raw data, online: \url{https://github.com/rsln-s/Importance-of-Kernel-Bandwidth-in-Quantum-Machine-Learning/}.

\section*{Acknowledgments}

We thank Ilya Safro and Sami Khairy for helpful discussions about classical kernel methods and feedback on the manuscript. We thank Evan Peters for sharing the preprocessed dataset used in ~\cite{peters2021machine}. This work was supported in part by the U.S.\ Department of Energy (DOE), Office of Science, Office of Advanced Scientific Computing Research AIDE-QC and FAR-QC projects and by the Argonne LDRD program under contract number DE-AC02-06CH11357. We gratefully acknowledge the computing resources provided on Bebop, a high-performance computing cluster operated by the Laboratory Computing Resource Center at Argonne National Laboratory.

\bibliography{sample}

\appendix 

\section{Additional numerical experiments}

\begin{figure*}[hb]
    \centering %
    \includegraphics[width=0.9\textwidth]{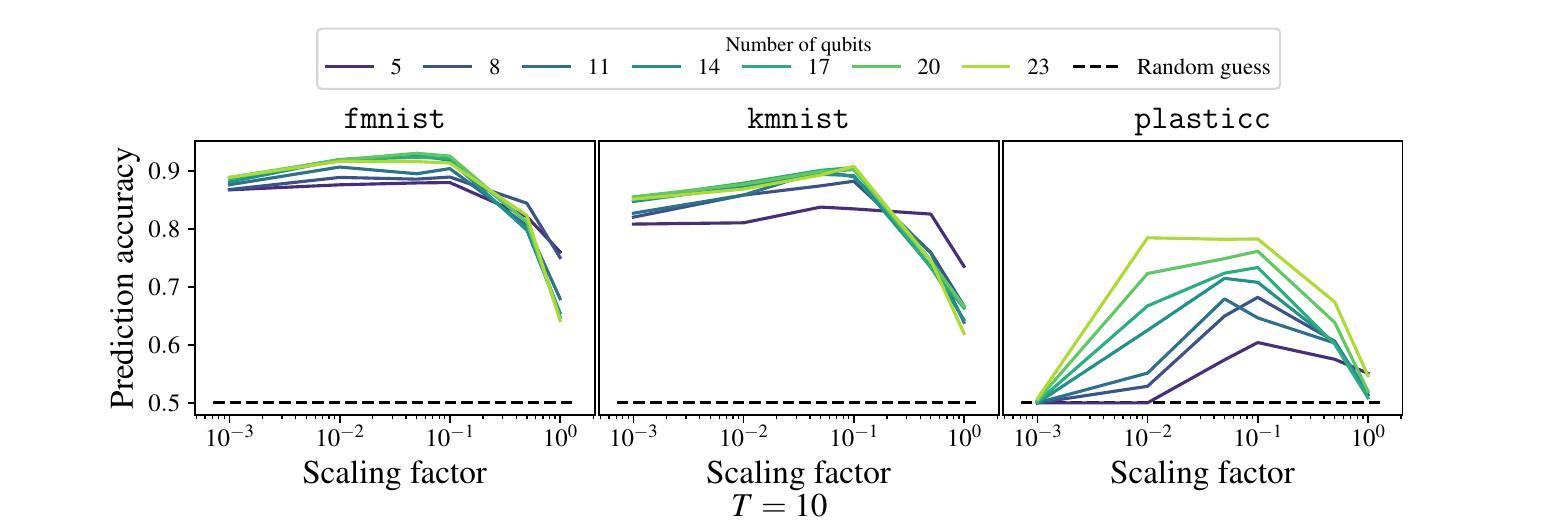}
    \includegraphics[width=0.9\textwidth, trim=0 0 0 0.5in]{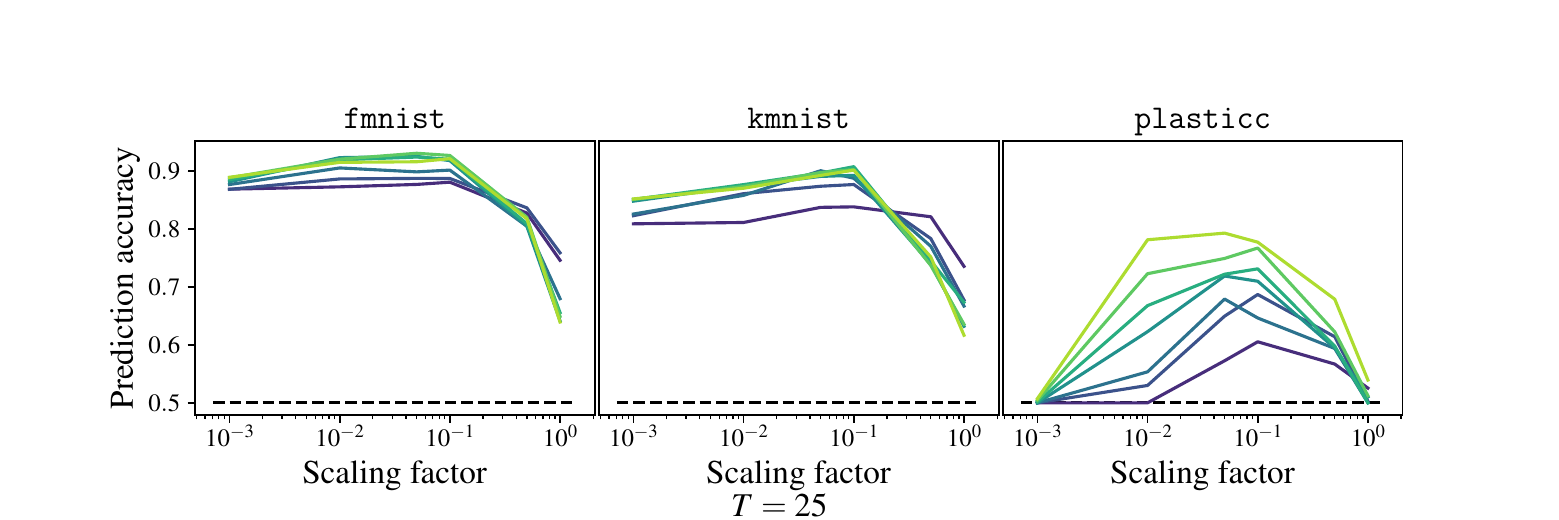}
    \caption{Prediction accuracy of the quantum kernel method with the Hamiltonian evolution feature map as a function of the scaling factor for the number of Trotter steps $T=25$ and $T=40$.}
    \label{fig:HamEvo_test_accuracy_extra_ntrotter}
\end{figure*}

\begin{figure*}
    \centering
    \subfloat[Prediction accuracy for the IQP feature map applied to the  \texttt{kmnist} dataset]{\includegraphics[width=0.19\textwidth, trim=0.8in 0 0.1in 0]{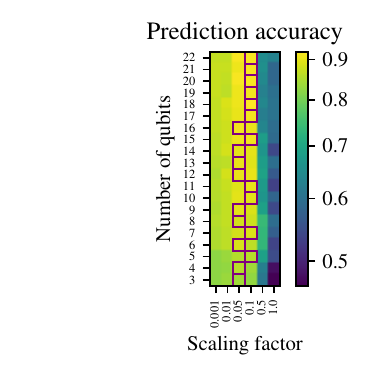}}
    \hspace{0.1in}
    \subfloat[Kernel matrix entries for the IQP feature map applied to the  \texttt{kmnist} dataset]{\includegraphics[width=0.19\textwidth, trim=0.8in 0 0.1in 0]{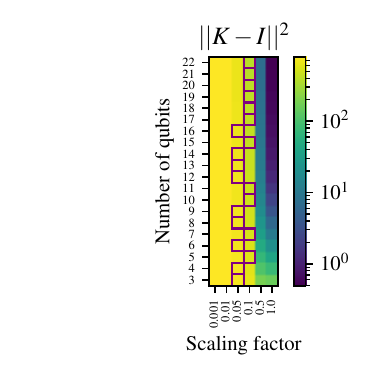}}
    \hspace{0.1in}
    \subfloat[Prediction accuracy for the IQP feature map applied to the  \texttt{plasticc} dataset]{\includegraphics[width=0.19\textwidth, trim=0.8in 0 0.1in 0]{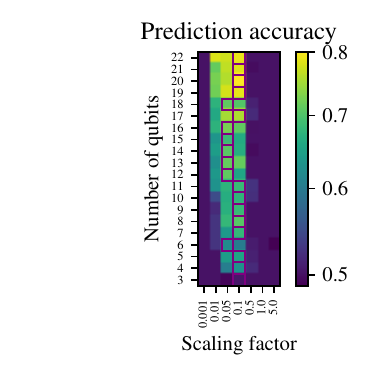}}
    \hspace{0.1in}
    \subfloat[Kernel matrix entries for the IQP feature map applied to the  \texttt{plasticc} dataset]{\includegraphics[width=0.19\textwidth, trim=0.8in 0 0.1in 0]{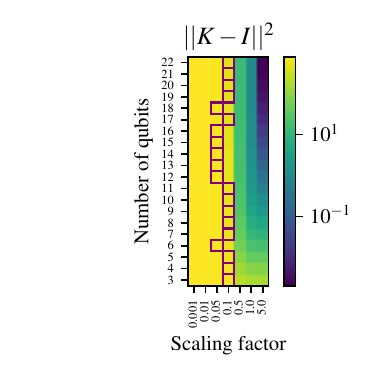}} \\
    \subfloat[Prediction accuracy for the Hamiltonian evolution feature map applied to the \texttt{fmnist} dataset]{\includegraphics[width=0.2\textwidth]{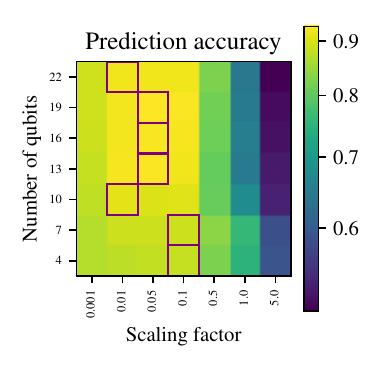}}
    \hspace{0.1in}
    \subfloat[Kernel matrix entries for the Hamiltonian evolution feature map applied to the \texttt{fmnist} dataset]{\includegraphics[width=0.2\textwidth]{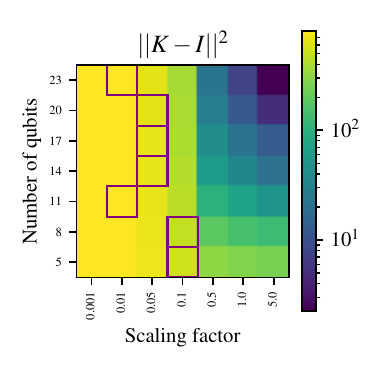}}
    \hspace{0.1in}
    \subfloat[Prediction accuracy for the Hamiltonian evolution feature map applied to the \texttt{kmnist} dataset]{\includegraphics[width=0.2\textwidth]{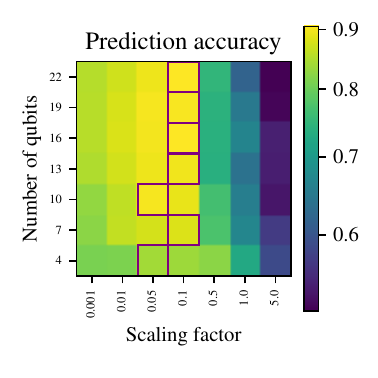}}
    \hspace{0.1in}
    \subfloat[Kernel matrix entries for the  Hamiltonian evolution feature map applied to the  \texttt{kmnist} dataset]{\includegraphics[width=0.2\textwidth]{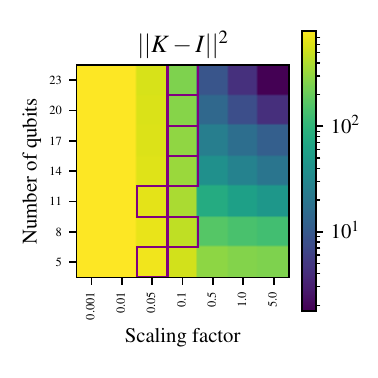}} \\
    \subfloat[Prediction accuracy for the Hamiltonian evolution feature map applied to the  \texttt{plasticc} dataset]{\includegraphics[width=0.2\textwidth]{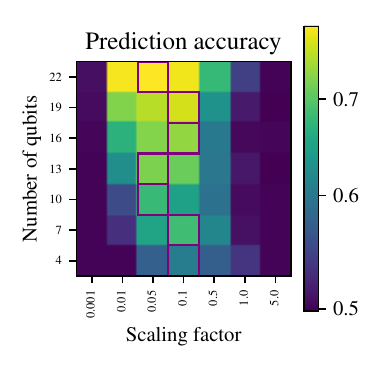}}
    \hspace{0.1in}
    \subfloat[Kernel matrix entries for the Hamiltonian evolution feature map applied to the \texttt{plasticc} dataset]{\includegraphics[width=0.2\textwidth]{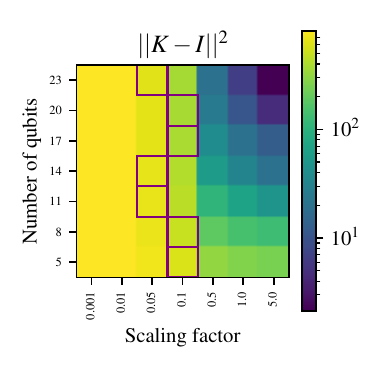}}
    \caption{Additional numerical evidence of kernel matrix entry scaling and model performance variation with the scaling factor. For each qubit count, the value of the scaling factor corresponding to the highest accuracy is highlighted with a purple square. For each model and each dataset considered, a clear ``Goldilocks'' regime is visible where the scaling factor is neither too small nor too large.}
    \label{fig:extra_2d}
\end{figure*}

Figure~\ref{fig:HamEvo_test_accuracy_extra_ntrotter} illustrates the relationship between model performance and scaling factor for varying levels of the second hyperparameter, namely, the number of Trotter steps $T$. No significant variation in performance with $T$ is observed. Figure~\ref{fig:extra_2d} shows the scaling of the kernel for feature maps and datasets not included in the main text.

\begin{figure*}[hb]
    \centering %
    \includegraphics[width=0.89\textwidth, trim=0 0 0 0.5in]{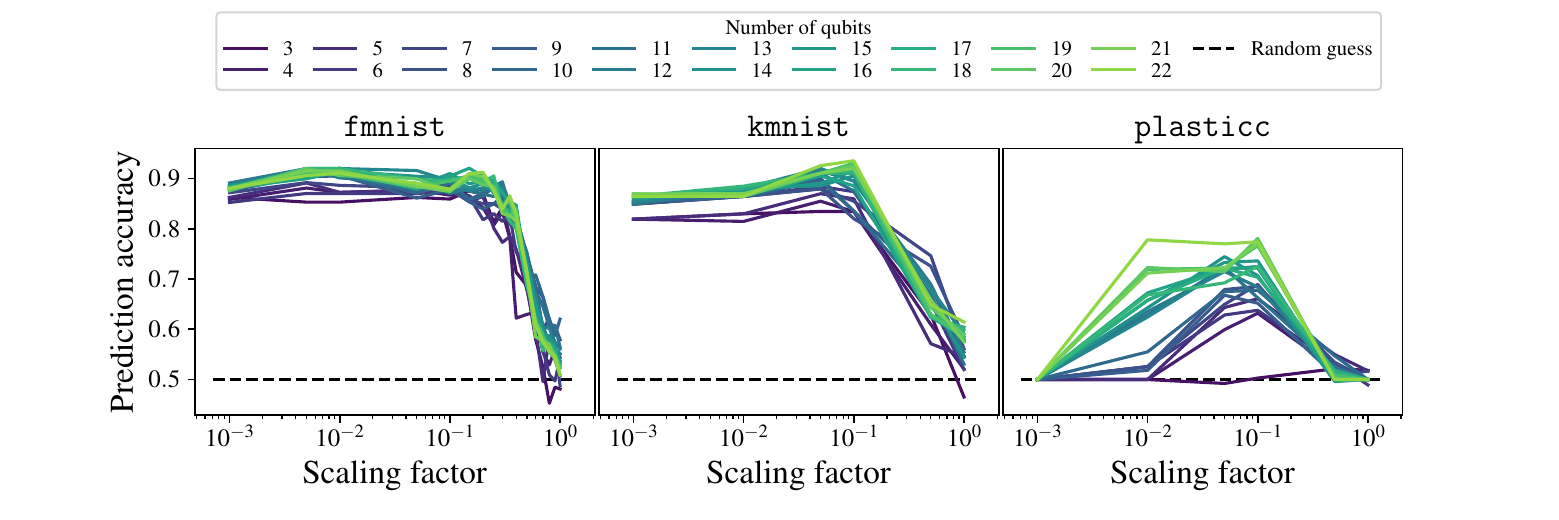}
    \includegraphics[width=0.9\textwidth, trim=0 0 0 0.5in]{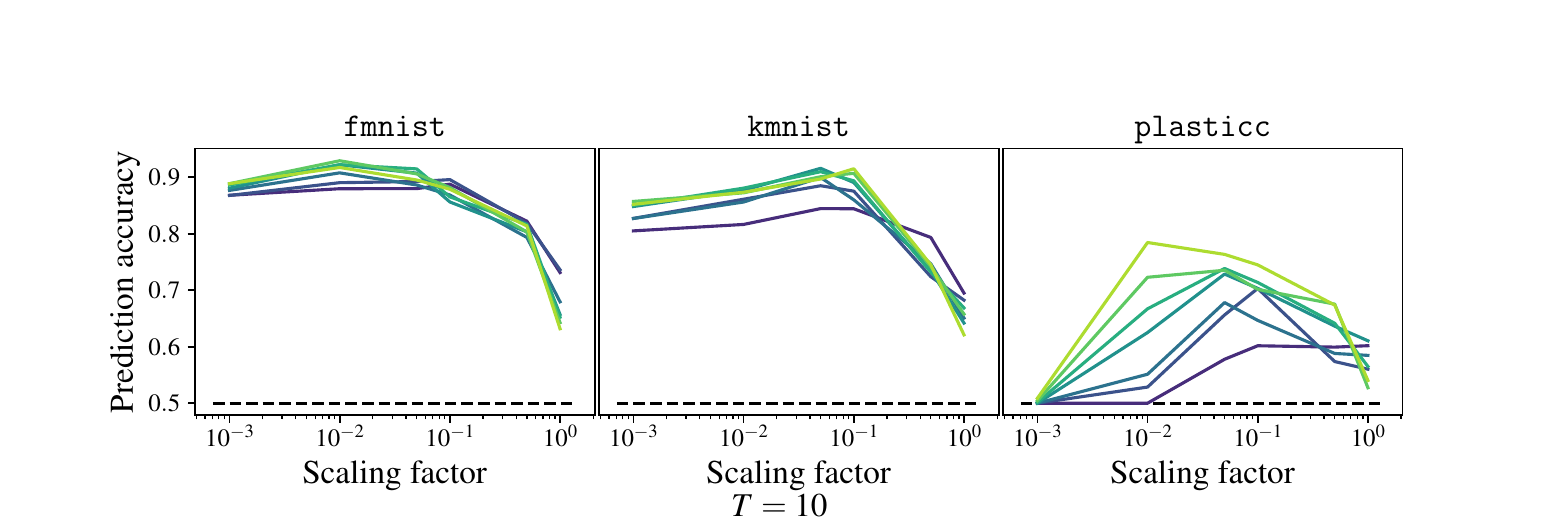}
    \includegraphics[width=0.9\textwidth, trim=0 0 0 0.5in]{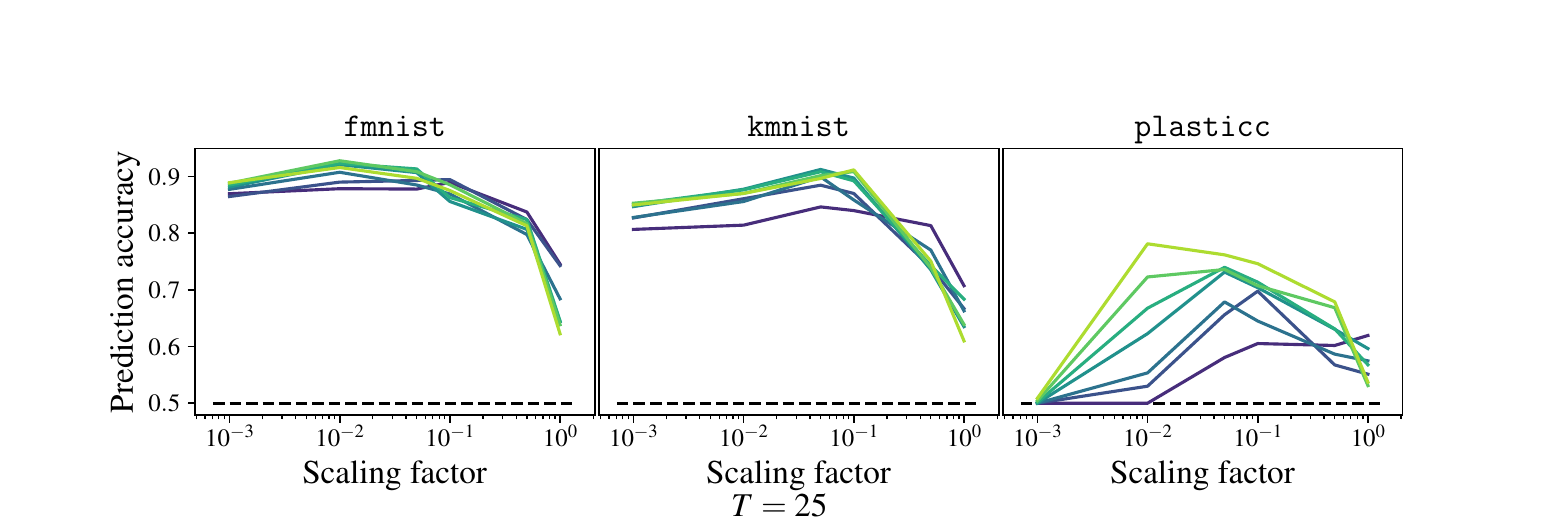}
    \includegraphics[width=0.9\textwidth, trim=0 0 0 0.5in]{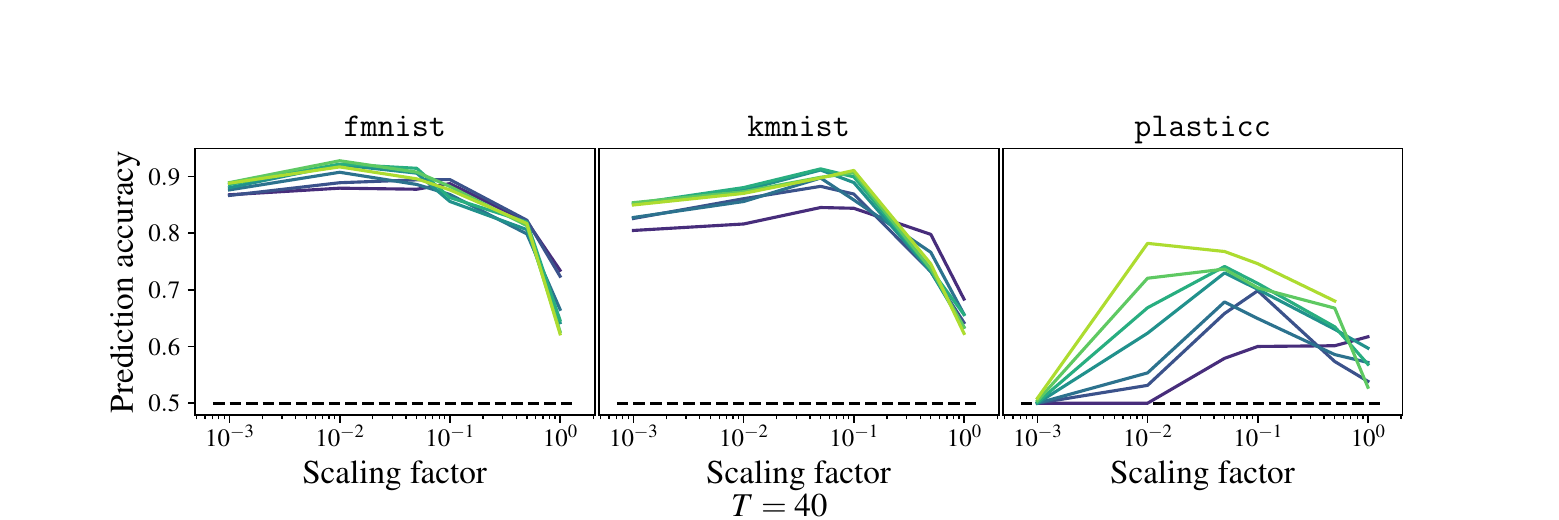}
    \caption{Prediction accuracy as a function of the scaling factor for IQP (top) and Hamiltonian evolution (bottom, varying $T$) without cross-validation for SVM hyperparameter optimization.}
    \label{fig:all_test_accuracy_dumb_CV}
\end{figure*}

\begin{figure*}
    \centering %
    \includegraphics[width=0.3\textwidth]{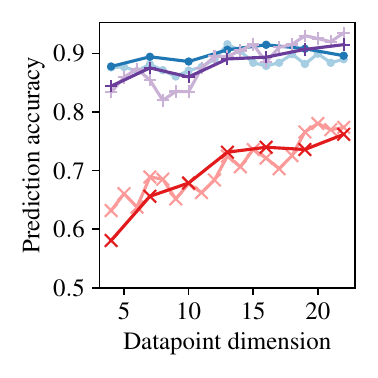}
    \caption{Prediction accuracy as a function of the datapoint dimension. Even without cross-validation, performance improves with qubit count.}
    \label{fig:all_over_under_fitting_dumb_CV}
\end{figure*}

Figures~\ref{fig:all_test_accuracy_dumb_CV}~and~\ref{fig:all_over_under_fitting_dumb_CV} show results without cross-validation for SVM hyperparameter optimization. Concretely, the penalty term $C$ is optimized to maximize the training score. Figure~\ref{fig:all_test_accuracy_dumb_CV} shows the prediction accuracy change with scaling factor, and Figure~\ref{fig:all_over_under_fitting_dumb_CV} shows scaling of prediction accuracy with qubit count for optimized bandwidth.

\section*{Disclaimer}
The opinions, findings, and conclusions or recommendations expressed in this material are those of the authors and do not necessarily reflect the views of JPMorgan Chase.

\end{document}